\documentstyle[12pt]{article}

\newcommand{\ez}{{\cal E}_0}
\newcommand{\seq}{Schr\"odinger equation$\ $}
\newcommand{\nn}{\nonumber}
\newcommand{\rf}{\ref}

\newcommand{\be}{\begin{equation}}
\newcommand{\ee}{\end{equation}}
\newcommand{\bea}{\begin{eqnarray}}
\newcommand{\eea}{\end{eqnarray}}
\newcommand{\r}{\rho}
\newcommand{\rai}{\rightarrow \infty}
\newcommand{\raz}{\rightarrow 0}

\newcommand{\for}{\quad \mbox{ for}\quad}
\newcommand{\aand}{\quad \mbox{ and}\quad}
\newcommand{\where}{\quad \mbox{ where}\quad}

\def\lb#1{\label{eq:#1}}
\def\rf#1{(\ref{eq:#1})}
\def\J{\mbox{\huge{J}}}
\newcommand{\G}{\Gamma}

\begin{document}

\baselineskip=14pt plus 0.2pt minus 0.2pt
\lineskip=14pt plus 0.2pt minus 0.2pt

\begin{flushright}
hepth@xxx/9408058 \\
 LA-UR-94-2569  \\
\end{flushright}

\begin{center}
\Large{\bf
 EXACT, ${\bf E=0}$,  SOLUTIONS FOR GENERAL
POWER-LAW  POTENTIALS. \\
II. QUANTUM WAVE FUNCTIONS} \\

\vspace{0.25in}

\large

\bigskip

Jamil Daboul\footnote{Email:  daboul@bguvms.bgu.ac.il}\\
{\it Physics Department, Ben Gurion University of the Negev\\
Beer Sheva, Israel}\\
$~~~~~~~$\\
and\\
$~~~~~~~$ \\
Michael Martin Nieto\footnote{Email:  mmn@pion.lanl.gov}\\
{\it
Theoretical Division, Los Alamos National Laboratory\\
University of California\\
Los Alamos, New Mexico 87545, U.S.A. }

\normalsize

\vspace{0.3in}

{ABSTRACT}

\end{center}
\begin{quotation}
For zero energy, $E=0$, we derive exact, quantum solutions for
{\it all} power-law potentials, $V(r) = -\gamma/r^{\nu}$, with
$\gamma > 0$ and $-\infty < \nu < \infty$.  The solutions are, in general,
Bessel functions of powers of $r$.  For $\nu > 2$ and $l \ge 1$
the solutions are normalizable; they correspond to states which are
bound by the angular-momentum barrier.
Surprisingly, the solutions for $\nu < -2$ are also normalizable,
They are discrete states but do not correspond to bound states.
For  $2> \nu \geq -2$  the states are unnormalizable continuum states.
The $\nu=2$  solutions are also unnormalizable, but are exceptional
solutions.  Finally, we find that
 by increasing the dimension of the \seq  beyond 4 an
effective centrifugal barrier is created, due solely to the extra
dimensions, which is enough to cause binding.   Thus, if $D>4$, there are
$E=0$ bound states for $\nu > 2$ even for $l=0$.
We discuss the physics of the above solutions are compare them to the
classical solutions of the preceding paper.

\vspace{0.25in}

\noindent PACS: 03.65.-w, 03.65.Ge

\end{quotation}

\vspace{0.3in}

\newpage

%***************************************************************************
%\baselineskip=.33in
%***************************************************************************

\section{Introduction}

In paper I \cite{db1}, we gave the general $E=0$, classical-orbit
solutions for all classical potentials of the form
\be
V(r) = -\frac{\gamma}{r^{\nu}}= -\frac{\gamma}{r^{2\mu+2}}~, ~~~~
\gamma>0~, ~~~~
-\infty<\nu,\mu<\infty~, \lb{pot}
\ee
where, as in the classical case, we will use for convenience
two parameters $\nu$ and $\mu$, which are related by
\be
\mu \equiv \frac{\nu -2}{2}, ~~~~~~~~ \nu = 2(\mu +1)~.
\ee
In this paper we will obtain the $E=0$ quantum solutions for the
power-law potentials \rf{pot}. We will find some remarkable
similarities between the classical and quantum solutions,
and demonstrate unusual quantum effects.

In Section 2 we derive the general solutions of the \seq for the
 potentials \rf{pot}, for all $\nu$ (except
$\nu =2$ or $\mu=0$, which is discussed separately in Section 5.2).
In Section 3 we find out which of these solutions are normalizable.
Then, in Section 4, we show that these normalizable states belong
to two classes, which have a different physical nature:
For $\nu >2$ or $\mu>0$, the normalizable solutions are
bound states if the angular-momenta satisfy $l>0$,
 just as the classical orbits were bound if the angular momentum was
not zero.  In contrast, the normalizable states
for $\nu < -2$ or $\mu < -2$, cannot be interpreted as bound states,
and has an unusual interpretation.

For  $-2 \leq \nu \leq 2$ or $-2 \leq \mu \leq 0$,
the solutions are not normalizable and hence correspond
to  free (unbounded) wave functions.
These are discussed in Section 5.
We discuss the solutions for special cases of $\nu$ in Section 6.
As a final result, in Section 7 we demonstrate a quantum mechanical
effect which is due to the dimension of the \seq .

In Section 8 we discuss the correspondence
between the classical solutions of paper I with the quantum
solutions.  This returns us to the starting point of paper I, the
``Folk Theorem" about the similar solvability or lack thereof of
equivalent classical and quantum systems. We close, in Section 9, with a
short summary.  (Two appendices deal with the choice of physical
solutions and with the $\gamma < 0$ problem.)

%***********************************************************

\section{General Quantum-Mechanical Solutions for All
$\nu \ne 2$ or $\mu \ne 0$}

The radial Schr\"odinger equation
with angular-momentum quantum-number, $l$, is
\be
E R_l =\left[-\frac{\hbar^2}{2m}\left(\frac{d^2}{dr^2}
+ \frac{2}{r}\frac{d}{dr}-\frac{l(l+1)}{r^2}\right)
+ V(r)\right] R_l ~.  \lb{scheq}
\ee
It is useful to express the \seq in terms of a dimensionless
radius
\be  \r \equiv r/a~,  \ee
where $a$ is a convenient quantity with the dimension of length.
Throughout this paper we also use a quantity with
the dimension of energy, given by
\be
   \ez \equiv \frac{\hbar^2}{2ma^2}~,
\ee
Dividing  Eq. \rf{scheq} by $\ez$ and using $\r$ yields
\be
 \frac{E}{\ez}R_l(\r)=\left[-\left(\frac{d^2}{d\r^2}
               + \frac{2}{\r}\frac{d}{d\r}-\frac{l(l+1)}{\r^2}\right)
          + \frac{ V(r)}{\ez}\right] R_l(\r)  ~. \lb{seq2}
\ee
If we express the potentials \rf{pot} in terms of $\r$, we have
\be
 V(r) = -~ \frac{\gamma}{r^\nu}
    \equiv - \ez \frac{g^2}{\r^\nu}~,  \lb{param}
 \ee
where $g^2$ manifestly is a dimensionless coupling constant.

Finally,  multiplying Eq. \rf{seq2} by $-\r^2$, setting $E=0$, and
substituting
the power-law potentials of Eq. \rf{param} into Eq. \rf{seq2}, we obtain
\be
0=\left[\r^2\frac{d^2}{d\r^2} + 2\r\frac{d}{d\r} - l(l+1)
         + \frac{g^2}{\r^{2\mu}}\right]R_l ~.
\lb{qmsp}
\ee

Eq. \rf{qmsp} is a form of the  differential equation for
a product of a power times a Bessel function of another power.
It can be written in the form \cite{mos}
\be
0=\left[z^2\frac{d^2}{dz^2} + (1-2\alpha)z\frac{d}{dz}
   +(\beta\eta z^{\eta})^2 +\left(\alpha^2 -\sigma^2\eta^2\right)
     \right] w(z)   \lb{qmb1}~,
\ee
with the general solution being \cite{mos1}:
\be
w=A~ z^{\alpha}J_{\sigma}(\beta z^{\eta})+B~ z^{\alpha}Y_{\sigma}(\beta
z^{\eta}),  \lb{qmb2}
\ee
where $A$ and $B$ are arbitrary constants. By considering the asymptotic
behavior of the two independent solutions, $J_{\sigma}(z)$ and
$Y_{\sigma}(z)$, we are led to choose the $J_{\sigma}(z)$ as the physical
solutions. This choice is explained in  Appendix A. Therefore, we have
\be
R_l(\r) = \frac{1}{\r^{1/2}}
           {\mbox{\huge{J}}}
           _{\left(\frac{l+1/2}{|\mu|}\right)}
           \left(\frac{g}
           {|\mu| \r^{\mu}}\right)
          = \frac{1}{\r^{1/2}}
           {\mbox{\huge{J}}}
           _{\left|\frac{2l+1}{\nu - 2}\right|}
           \left(\frac{2g}
           {|\nu - 2|\r^{\left(\frac{\nu - 2}{2}\right)}}
           \right)~, ~~~ \mu \neq 0~.  \lb{gensol}
\ee
Note, in particular, that the $\mu$ that
labels the power of $\r$ in the argument
of $J$ does {\it not} have an absolute value.

It is interesting to observe that for special indices, $\mu$,
the radial wave functions \rf{gensol} become proportional to
spherical Bessel functions
\be
j_n(z) = \sqrt{\frac{\pi}{2z}}J_{n+1/2}(z)
 = -z^n\left(-\frac{1}{z}\frac{d}{dz}\right)^n \frac{\sin z}{z} ~.
\label{sphereJ}
\ee
This happens for a specific $l$
if $|\mu|$ can be written as a ratio of the form
\be
|\mu| =\frac{l+1/2}{n+1/2}=\frac{2l+1}{2n+1}~.
\lb{con1}
\ee
In particular,  for the specific index, $\mu=1$,
the condition \rf{con1} is always true for all $l$.
(The example
discussed in Section 4.2 corresponds to $\mu=-3$, $l=1$ and $n=0$.)

%******************************************************

\section{Normalizable States}

The normalization
constants for these wave functions, if they exist, would be equal to
\be
N_l^{-2} = \int_{0}^{\infty}\frac{r^2dr}{\r}
             {\mbox{\huge{J}}}^2
           _{\left(\frac{l+1/2}{|\mu|}\right)}
           \left(\frac{g}
           {|\mu| \r^{\mu}}\right)~, ~~~~ \mu \ne 0. \lb{norm}
\ee
Changing variables from $r$ to $\r$ and then from $\r$ to
$t=\r(|\mu|/g)^{1/\mu}$, one obtains
\be
N_l^{-2} ={a^3}\left(\frac{g}{\mu}\right)^{2/\mu}~
        \left[-~\frac{1}{\mu}
      \int_{0}^{\infty}{t~dt~}
             {\mbox{\huge{J}}}^2
           _{\left(\frac{l+1/2}{|\mu|}\right)}
           \left(\frac{1}
           {t^{\mu}}\right)\right]~.
\ee

We now want to change variables to $z=t^{-\mu}$.  In doing so, we must be
careful when converting the limits of integration.  For $\mu>0$,
the change
inverts the limits of integration, introducing another minus sign.  So,
the integral's prefactor  $-1/\mu \rightarrow +1/\mu=1/|\mu|$.
Contrariwise, for $\mu<0$ the limits stay the same.  But in this case, we
have,  $-1/\mu=1/|\mu|$.  Therefore, in all cases we have
\be
N_l^{-2} = a^3\left(\frac{g}{|\mu|}\right)^{2/\mu}~
        \frac{I_l}{|\mu|}~,
\ee
where
\be
I_l = \int_0^{\infty}\frac{dz}{z^{(1+2/\mu)}}
              {\mbox{\huge{J}}}^2
           _{\left(\frac{l+1/2}{|\mu|}\right)}(z) ~.  \lb{int1}
\ee

The integral \rf{int1} is a very special case
of the difficult Weber-Schafheitlin discontinuous integral, whose
integrand involves a product of two Bessel functions of different
orders and different arguments. The generral integral is not
continuous as a function of the arguments of the Bessel functions as
they approach each other.
Details may be found in Watson's book \cite{wat}.  The end
result is of the form  \cite{wat2,tran,mos2,gr}
\bea
I_l(\sigma,\lambda) &=& \int_0^{\infty}\frac{dz}{z^\lambda}
              J^2_\sigma (z) =
\frac{\Gamma(\lambda)}{2^\lambda \Gamma^2\left(\frac{1+\lambda}{2}\right)}
\frac{\Gamma\left(\frac{1-\lambda+2\sigma}{2}\right)}
{\Gamma\left(\frac{1+\lambda+2\sigma}{2}\right)}  \nn \\
&=& \frac{\Gamma(\lambda/2)}{2 \pi^{1/2}
\Gamma\left(\frac{1+\lambda}{2}\right)}
\frac{\Gamma\left(\frac{1-\lambda}{2}+\sigma\right)}
{\Gamma\left(\frac{1+\lambda}{2}+\sigma\right)}~,
 \lb{ig}
\eea
valid  provided that the following two conditions
are satisfied:
\be
1+ 2 Re(\sigma)~ >~ Re(\lambda)~ >~0~.
\ee
[The last expression in Eq. \rf{ig} is obtained \cite{mos3}
by using the duplication formula
\be
\G(2x)=\G(x)~\G(x+1/2)~2^{2x-1}~\pi^{-1/2}
\ee
to factorize $\G(\lambda)$.]

Applying Eq. \rf{ig} to the integral \rf{int1}  we obtain
\be
I_l= \frac{\Gamma\left(\frac{1}{2}+\frac{1}{\mu}\right)}
{2\pi^{1/2} \Gamma\left(1+\frac{1}{\mu}\right)}
\frac{\Gamma\left(\frac{l+1/2}{|\mu|}-\frac{1}{\mu} \right)}
{\Gamma\left(1+\frac{l+1/2}{|\mu|}+\frac{1}{\mu}\right)}~,
\ee
which is finite, if
\be
1+ \frac{2l+1}{|\mu|} ~>~ 1+\frac{2}{\mu} ~>~ 0~.   \lb{cond}
\ee

%*********************************

\section{Classes of Normalizable States}

\subsection{Bound states at the scattering threshold: $\nu >2$ or
$\mu>0$ with $l \geq 1$}

The conditions \rf{cond} are satisfied for
 $\mu>0$ or $\nu >2$ and  $l>1/2$.  These are normal bound states, and
exactly correspond to the conditions
for  classically bound orbits described
in paper I.  In the classical problem, there were $E=0$ bound states for
any non-zero angular momentum with $\nu>2$.  In quantum mechanics, the
smallest non-zero angular momentum allowed is $l=1>1/2$.

This physics is not what one normally expects
in quantum mechanics.  Usually one thinks
that the $E=0$ state at the edge of the continuum is free.  That intuition
comes from thinking of the
Coulomb system, where the potential asymptotes to zero from below at
$r\rightarrow\infty$.   (See Fig. 10 of Paper I.)

Here, the potential asymptotes to zero from above, as in Figure 1 of Paper I.
Thus, for $E=0$ the wave function
which is concentrated at the origin must tunnel to infinity to reach another
free region.
That takes forever, and so the state is bound.

\subsection{Normalizable yet ``unbound" states: $\nu <- 2$ or $\mu<0$}

We now come to an unusual set of normalizable states.  The conditions
\rf{cond} are satisfied for all $\nu < -2$ or  $\mu<-2$, for any $l$.
(Technically, for all $l>-1/2$.)
Since in this case the potentials are strongly repulsive, this
result  at first seems counter-intuitive.

Note that for $\nu <2$  the power potentials \rf{pot} are {\it more}
repulsive than the inverse harmonic oscillator.   The unusual spectral
properties of such potentials are known to the mathematical physics
community.
What we have found here is an explicit  solution for $E=0$ and
all potentials \rf{pot} with $ \nu <-2$.

It can be shown, using asymptotic argumntes, that normalizable solutions
also exact for  the
 continuum spectrum $E\ne 0$. However, by imposing proper
boundary conditions, a {\em discrete} subset can be chosen,
which  makes  the
Hamiltonian self adjoint.\cite{klau,berez,wightman,case}.

Physically, one can see a classical analogue
 for this unusual behavior.  Recall
the result given at the end of Sec. 4.4 of paper I.  We noted that
for $\nu < -2$ or  $\mu<-2$ it took a {\it finite} time for the
particle to go from its turning point, $a$, to infinity.  Thus, although
these states are ``unbound," the proportion of time they spend outside
a sphere of finite radius $R$ can be made as small as desired.

Classically, when  proper boundary conditions are imposed,
the  solutions are periodic in time \cite{dn3} in the sense that
the particle
goes back and forth between infinity and the turning
point, $a$, with definite time intervals.
Such  classical orbits correspond  to
the normalizable, discrete quantum solutions.

It can be useful to the reader to look at a simple, special case.
If one considers $\mu= -3$ and $l=1$, the solution is a spherical
Bessel function with one term, and hence easily normalized.

%*********************************************************

\section{Unnormalizable Solutions}

\subsection{Normal bound states:  $\nu > 2$ or $\mu > 0$ with $l \geq 1$}

The unnormalizable solutions for the potentials \rf{pot} fall into three
categories.  The first two are
normal,  unbound states.

i) ~The S-wave ($l = 0$) solutions of the attractive potentials,
with $\nu>2 $ or $\mu>0$.

Here, there is no forbidden tunneling region
to prevent the wave function from traveling  to infinity.

ii) ~The
 solutions of the potentials, with  $-2 \leq \nu < 2$ or
$-2 \leq \mu < 0$ and all $l$.

In this case, the corresponding effective
potentials, $ U(\r)=\ez(l(l+1)/\r^2-g^2/\r^\nu $, further split into
two types.  The potentials are
Kepler like, for $ 0< \nu <2$, in that they have a minimum for $l\ne 0$
at a finite $\r$.  Otherwise,  for $ -2< \nu \le 0$, the potentials
 are ``weakly" repulsive, meaning
the classical particle takes an infinite time to reach infinity.

\subsection{The exceptional ``free" solution: $\nu = 2$ or $\mu = 0$}

iii) The exceptional power $\nu=2$.

Just as in the classical theory, the $\nu = 2$ or $\mu = 0$ case
is the one exceptional case which does not fit into the general solution,
which is given here in Eq. \rf{gensol}.
This is the quantum-mechanical analogue of having
an inverse-square  potential $V(r)=-\gamma/r^2$, that cannot be
differentiated from the angular-momentum barrier.

The value of $g^2$ relative to $l(l+1)$ leads to three kinds of
effective potentials, $U(\r)$: purely positive, identically zero,
and purely negative.
In quantum mechanics there
are zero-energy solutions for each of the above 3 cases.
(Contrariwise, since the kinetic energy is  classically nonnegative,
there are nontrivial $E=0$ solutions only for negative effective potentials,
$U <0 $.)

The Schr\"odinger equation,
\be 0=\left[\frac{d^2}{d\r^2}
+ \frac{2}{\r}\frac{d}{d\r}
+ \frac{ g^2-l(l+1)}{\r^2}\right] R_l~,
\lb{seq} \ee
is homogeneous in ${d}/{d\r}$ and ${1}/{\r}$. Therefore,  its
solutions are expected to be powers of $\r$:
\be
 \left[\r^2 \frac{d^2}{d\r^2}
+ 2\r\frac{d}{d\r}+ g^2-l(l+1)\right] \r^\alpha
= [ \alpha^2 +\alpha + g^2-l(l+1) ] \r^\alpha =0~.
\ee
The two roots of the above quadratic equation are
\be
\alpha_\pm = - \frac{1}{2}\pm \sqrt{ (l+1/2)^2 - g^2} =
 - \frac{1}{2}\pm \sqrt{\Delta}~,
\ee
where
\be \Delta = (l+1/2)^2 - g^2~.
\ee
Depending on the value of $\Delta$, we obtain three types of solutions
of Eq. \rf{seq}:
\be
R_l =\left\{ \begin{array}{ll}
        \r^{-1/2}[A~ \r^{\sqrt{\Delta}} +
                              B~   \r^{-\sqrt{\Delta}}]
                     &  \mbox{ $ l+1/2 > g $} \\
                 ~~~ & ~~~ \\
              \r^{-1/2}[A~ +  B~ \ln{\r}]
                     &  \mbox{ $ l+1/2 = g$}  \\
                 ~~~ & ~~~ \\
              \r^{-1/2}[A~ \cos (|\Delta|^{1/2} \ln{\r})
                      +  B~ \sin (|\Delta|^{1/2} \ln{\r})]
                     &  \mbox{ $ l+1/2 < g$}
         \end{array}   \right. ~, \lb{solp}
\ee
where $A$ and $B$ are arbitrary constants.

The  solutions in the first line of Eq. \rf{solp} hold when
$\Delta > 0$.  They correspond to
powers,
whose orders are the two real square roots of $\Delta$, which have
opposite signs.
These square roots approach each other as
$\Delta \rightarrow 0$.   Then,
as $\sqrt{\Delta}\rightarrow +0$ and
$\sqrt{\Delta}\rightarrow -0$, the solutions become
the constant and logarithmic terms, respectively, given on the second line
of Eq. \rf{solp}.

Finally, for  $\Delta<0 $, the square roots become pure imaginary,
and thus leads to trigonometric behavior.

Note that Eq.
\rf{seq} for $g=0$ corresponds to a free particle with zero energy.
For this case, the solutions with $ l+1/2 > g = 0$ in \rf{solp} are
relevant.   They then yield
\be
R_l= A~ r^l + B~ r^{-(l+1)}~,
\ee
 as expected.

%*********************************************************

\section{Special Quantum-Mechanical Cases}

\subsection{$\nu = 4$ or $\mu = 1$}

The potential $V(r)=-\gamma/r^4$ classically leads to the bound orbit
which is a circle.  The quantum solution Eq. \rf{gensol}
simplifies to
\be
R_l = \r^{-1}~j_l({g}{\r}^{-1})~,
\ee
where  $j_n$ is the spherical Bessel function defined in Eq.
(\ref{sphereJ}). It corresponds to a bound states, for $l\ge1$.

\subsection{Coulomb potential: $\nu = 1$ or $\mu = -1/2$}

This is the zero-energy solution for the Coulomb problem.  From Eq.
\rf{gensol} above, we have
\be
R_l = \r^{-1/2}~ J_{2l+1}(2g\r^{1/2}) ~.
\ee
These are continuum unnormalizable states.

\subsection{Constant potential: $\nu = 0$ or $\mu =-1$}

This corresponds to a constant negative potential of depth $g^2$ below
the angular-momentum barrier.  So, its solution is similar to that of
a free particle with angular-momentum $l$ and positive total energy
$g^2 \ez = (\hbar k)^2/2m $, where $k=p/\hbar=g/a$ is the
effective wave number.
The solution is  a spherical Bessel function,
\be
R_l = j_l(g\r)=j_l(kr)~,\where k=g/a~.
\ee

\subsection{Inverted harmonic-oscillator potential: $\nu =\mu= -2$}

Our final special case is that of a repulsive (negative)
harmonic-oscillator potential, $V=-\gamma r^2$.
The solution is
\be
R_l = \r^{-1/2} ~J_{(l/2+1/4)}(g\r^2/2)~.
\ee

%************************************************************

\section{Bound States in Arbitrary Dimensions}

One can easily generalize the problem of the last
section to arbitrary $D$
space  dimensions.   Doing so yields another surprising
physical result.

To obtain the
$D$-dimensional analogue of Eq.\rf{qmsp},  one simply has to
replace $2\r$ by $(D-1)\r$ and $l(l+1)$ by $l(l+D-2)$ \cite{nd}:
\be
0=\left[\r^2\frac{d^2}{d\r^2} + (D-1)\r\frac{d}{d\r} - l(l+D-2)
        +\frac{g^2}{\r^{2\mu}}\right]R_{l,D}~.
\lb{qmD}
\ee
This equation is again a generalized Bessel equation. Comparing
\rf{qmD} with \rf{qmb1}, yields the following physical solutions:
\bea
R_{l,D} &=&\frac{1}{\r^{D/2-1}}
           {\mbox{\huge{J}}}
           _{\left(\frac{l+D/2-1}{|\mu|}\right)}
           \left(\frac{g}
           {|\mu| \r^{\mu}}\right)
                        \\ \nonumber
     &=&
 \frac{1}{\r^{D/2-1}}
           {\mbox{\huge{J}}}
           _{\left(\frac{2l+D-2}{|\nu - 2|}\right)}
           \left(\frac{2g}
           {|\nu - 2|\r^{\left(\frac{\nu - 2}{2}\right)}}
           \right)~.
\eea

To find out which states are normalizable one first has to change the
integration measure  from $r^2dr$ to $r^{D-1}dr$ and again
proceed as before. The end
result is that if the wave functions are normalizable,
the normalization constant is given by
\be
N_{l,D}^{-2} =\frac{a^D}{|\mu|}\left(\frac{g}{|\mu|}\right)^{2/\mu}~
I_{l,D}~~,
\ee
where
\be
I_{l,D} =  \int_0^{\infty}\frac{dz}{z^{(1+2/\mu)}}
              {\mbox{\huge{J}}}^2
           _{\left(\frac{l+D/2-1}{|\mu|}\right)}
           (z) ~. \lb{ild}
\ee
We see that the above integral is exactly
equal  to that in Eq. \rf{int1},
except that $l$ is replaced by the effective quantum number
\be
l_{eff}=l+\frac{D-3}{2}~. \lb{leff}
\ee
Therefore,
\be
I_{l,D} = \frac{1}{2\pi^{1/2}}
\frac{\Gamma\left(\frac{1}{2}+\frac{1}{\mu}\right)}
{\Gamma\left(1+\frac{1}{\mu}\right)}
\frac{\Gamma\left(\frac{l+D/2-1}{|\mu|}-\frac{1}{\mu}\right)}
{\Gamma\left(1+\frac{l+D/2-1}{|\mu|}+\frac{1}{\mu}\right)}~,
\ee
which is defined and convergent for
\be
\frac{2l+D-2}{|\mu|}+1 >\frac{2}{\mu}+1>0~.
 \lb{gtD}
\ee

This yields the surprising result that there are bound states for
all $\nu > 2$ or $\mu > 0$ when $l >2-D/2$.  Explicitly this
means that the minimum allowed $l$ for there to be zero-energy
bound states are:
\bea
D = 2~, ~~~~~ l_{min} &=& 2~, \nonumber \\
D = 3~, ~~~~~ l_{min} &=& 1~, \nonumber \\
D = 4~, ~~~~~ l_{min} &=& 1~, \nonumber \\
D > 4~, ~~~~~ l_{min} &=& 0~. \lb{lmin}
\eea

This effect of dimensions is purely quantum mechanical and can be
understood as follows:  Classically, the number of dimensions involved
in a central potential problem has no intrinsic effect on the
dynamics.  The orbit remains in two dimensions, and the problem is
decided by the form of the effective potential, U, which contains
only the angular momentum barrier and the dynamical potential.

In quantum mechanics there are actually two places where
an effect of dimension appears.  The first is in the factor $l(l+D-2)$
of the angular-momentum
barrier.  The second is more
fundamental. It is due to the operator
\be
U_{qm} = -\frac{(D-1)}{\r}\frac{d}{d\r}~.
\ee
The contribution of $U_{qm}$ to the ``effective potential"
can be calculated by using the ansatz
\be R_{l,D}(\r)\equiv \frac{1}{\r^{(D-1)/2}}\; \chi_{l,D}(\r)~. \lb{ansatz}
\ee
This transforms the $D-$dimensional radial \seq into a
$1-$dimensional Schr\"odinger equation in the $\rho$ variable:
\be
0=\left[-\frac{d^2}{d\r^2}+ U_{l,D}(\r)\right] \chi_{l,D}~.
\lb{qmDeff}
\ee
In Eq. \rf{qmDeff}, the effective potential $U_{l,D}(\r)$ is given by
\bea
U_{l,D}(\r)&=&\frac{(D-1)(D-3)}{4 \r^2} + \frac{l(l+D-2)}{\r^2}+
V(\r)\nn \\
&=&
\frac{l_{eff}(l_{eff}+1)}{\r^2}+V(\r)~,
\lb{udeff}
\eea
with $l_{eff}$ given in Eq. \rf{leff}. Since the \seq \rf{qmDeff} depends
only on the combination
$l_{eff}$, the solution $\chi_{l,D}(\r)$ does not depend
on $l$ and $D$ separately.
This explains, in particular, the values of $l_{min}$ given in
Eq.  \rf{lmin}.

Although the above ansatz is well known, the
dimensional effect has apparently not been adequately appreciated.
One reason may be attributed to the fact that in going from $D=3$ to
$D=1$, $l_{eff}$ remains equal to $l$.

However, in our problem this effect leads to a physical result,
which is so counter-intuitive, that it cannot be overlooked.

The dimensional effect essentially produces an additional
centrifugal barrier
which can bind the
wave function at the threshold, even though the
expectation value of the angular momentum vanishes.  Note that this
is in distinction to the classical problem, where there would be no
``effective" centrifugal barrier to prevent the particle from
approaching $r \rightarrow \infty$.

%**************************************************

\section{Classical vs. Quantum: The ``Folk Theorem"}

We found here, exactly as in  classical
physics \cite{db1}, that except for the $\nu = 2$ case,
the $E=0$ radial solutions of the Schr\"odinger
equation \rf{qmsp}
are  given in terms of a single function, presented in Eq.   \rf{gensol}.
Although the functional form of the solutions
is the same for all $\nu \ne 2$, the properties of these solutions
change drastically as one passes over three regions of the index, $\nu$.

(1) For $ \nu >2$, the radial wave functions are normalizable
if $l > 1/2 > 0$. These are bound states,  analogous to the classical
bound orbits.
 The intuitive explanation for the existence of these  bound states
is that the tunneling distance is infinite.
The forbidden region  itself is due to the
repulsive angular-momentum barrier.
(For $l=0$ there is no such infinite barrier, and so the solutions
are free, as are
the  $-2 \leq \nu < 2$ cases described below.)

Observe that here, as in the classical case, the virial theorem is
violated.  Neither $\langle V \rangle$ nor $\langle -\nabla^2 \rangle$
is finite.

(2) For $-2 \leq \nu \le 2$ and all $l >0$
the effective potential,
$U(r)$, is repulsive near the origin, because in this region
the centripetal potential
dominates. This classically insures that  the
particle  does not approach the origin
beyond a certain minimal distance,  $r=a$.  Further, $U(\r)$ for
$\r \rightarrow \infty$
approaches zero from below.  This means that
the wave functions are unnormalizable and are part of the continuum.

(3) For $\nu < -2$,  the unusual quantum solutions mimic the
unusual classical solutions.  Classically, the travel time to infinity
is finite! This raises an interesting question: {\em
What happens classically to the particle after it reaches infinity?}

The classical answer is that the solutions are periodic in time.
Quantum mechanically this corresponds
to imposing boundary conditions on the wave functions.
These boundary conditions imposed on the normalizable solutions
yield a discrete spectrum instead of a continuous one. In this way
one obtains a proper self adjoint extension of the
Hamiltonian, and the unitarity of the transition
 operator $ U(t)= exp[-it H]$ is thus assured \cite{klau,berez,case,wightman}.
The special potential with $\nu = 2$ has exceptional solutions,
 both in  classical physics and also in  quantum physics.

Therefore, we see that
the mathematical similarities between
all the classical and quantum-mechanical
solutions are intriguing.  So, too, are the physical meanings of the
solutions.  How true, one must ask, is the ``Folk Theorem?"

%***********************************************************

\section{Summary}

The reason  the $E=0$ solutions
could be solved analytically was because there is one less
$\r$-power-term in the  differential equations of Newton and Schr\"odinger
which must be dealt with.
This  enabled us to obtain exact results in both cases.
We could then study the properties of the solutions explicitly and make
concrete comparisons between the classical and the quantum cases.
Sometimes the classical solutions
were intuitively helpful in understanding the quantum
solutions, and sometimes the opposite was true.
In any event, the
interest in the ``Folk Theorem" was amply rewarding.

Along the way there were unusual mathematical problems, due to the
singular behavior of the potentials \rf{pot}, which stood in the
way of physical understanding.
For example, instead of the standard requirement that $R(0)$ be
finite at the origin, we only demanded the physically reasonable
condition, that the probability of finding the particle near the origin
should be finite. In this way we could accept, as
physical solutions,  functions which are not
analytic at $\r=0$. (See appendix A.3)

The study of the quantum solutions led us to demonstrate three
interesting effects.

 (1) There exist  permanent
bound states at the threshold, {\em for all $l>0$ and all $\gamma >0$}.
That is,  there are $E=0$ states which  persist if we change
the coupling constant $\gamma$  by a positive factor.

In contrast,  $E=0$ bound states which exist in the literature
\cite{long,schiff2,do,barut} occur only ``accidentally,"  i.e.,
 for  special values
of the coupling constants. These states occur when the
bound-state pole in the partial-wave amplitude is just crossing the
scattering threshold to reach the second sheet.
Thus,  they become a decaying resonance.
Such an accidental crossing of the threshold does not occur for all $l>0$
simultaneously. A review of these accidental bound states is given in
\cite{dn3}.

 (2) There exist  normalizable solutions for
highly repulsive potentials ($\nu<-2$).

 (3) For higher-space dimensions, each additional
dimension adds a half unit to the effective angular-momentum
quantum number, $l_{eff}$, of Eq. \rf{leff}.
An  effective centripetal barrier,  solely due to this dimensional
effect,  i.e., even for $L^2=0$,  is capable of producing a bound state
when $D>4$.
This result is a  remarkable manifestation of quantum mechanics \cite{ND}
and  has no classical counterpart.

\newpage

%*****************************************************************

\section*{Appendix A: Asymptotics and Choice of Physical Solutions }

\subsection*{A.1: Asymptotics of Bessel functions}

The physical solutions of the radial \seq  must be linear combinations
of the solutions of the generalized Bessel equation \rf{qmb1}.
To determine these combinations, we first consider the behavior of
the wave functions at the origin and for large radii, $\r$.

Recall the asymptotic behavior of the Bessel functions:
\be
J_\sigma(z)\sim z^\sigma~, ~~~~ Y_\sigma(z)\sim z^{-\sigma}~,
{}~~~~ z\rightarrow 0~,  \lb{asz}
\ee
and
\be
J_\sigma(z)\sim \frac{1}{\sqrt{z}} \cos(z-c_\sigma)~,
{}~~~~ Y_\sigma(z)\sim  \frac{1}{\sqrt{z}} \cos(z-\tilde {c}_\sigma)~,
{}~~~~ z\rightarrow \infty~,  \lb{asi}
\ee
where the $c_\sigma$ and $\tilde{c}_\sigma$ are constant phases, which
depend on the index $\sigma$.
The argument  of the Bessel functions in
the solutions \rf{gensol} depend on $\r$ as
\be
z= \frac{g}{|\mu| \r^\mu}~, ~~~~ \mu\ne 0~.
\ee
Therefore, the $z\raz$ limit of Eq. \rf{asz} gives
\bea
R_l(\r)&=&\frac{1}{\sqrt{\r}}\J_{\frac{l+1/2}{|\mu|}}(g/|\mu|\r^\mu)\sim
\frac{1}{\sqrt{\r}}(\r^{-\mu})^{\frac{l+1/2}{|\mu|}}=
\frac{1}{\sqrt{\r}}\r^{-(l+1/2) \mbox{sgn}\, \mu} \lb{asm4}\\
&=& \left\{ \begin{array}{lll}
         \r^{-(l+1)}      \rightarrow  0 & \for \r \rightarrow \infty
&\aand \mu~>~0~, \\
\r^{l}      \rightarrow 0 & \for \r \rightarrow 0 &\aand \mu~<~ 0 ~.
                  \end{array}   \right.  \lb{solp2}
\eea
Similarly, the $z\rai$ limit of Eq. \rf{asi} yields the following upper
bounds:
\bea
|R_l(\r)|&=&\left| \frac{1}{\sqrt{\r}}
\J_{\frac{l+1/2}{|\mu|}}\left(\frac{g}{|\mu|\r^\mu}\right) \right|\sim
\left|
\r^{\frac{\mu-1}{2}} \cos\left(\frac{g}{|\mu|\r^{\mu}}-const.\right)\right|
 \le
\r^{\frac{\mu-1}{2}}  \lb{asm5}\\
&= &\left\{ \begin{array}{lll}
        \r^{(|\mu|-1)/2}      \rightarrow   0 & \for \r \raz &\aand \mu~>~1~,
\\
    \r^{-(1-\mu)/2}      \rightarrow \infty & \for \r \raz &\aand 1>\mu>0~, \\
         \r^{-(|\mu|+1)/2}    \rightarrow  0 & \for \r \rai &\aand \mu~<~ 0 ~.
                  \end{array}   \right.  \lb{solp3}
\eea

\subsection*{A.2: The physical solutions for $\mu < 0$}

\noindent From Eq. \rf{solp2} it follows that
the solutions \rf{gensol} for $\mu <0$
behave as $R_l(r) \sim \r^l$, for $\r \raz$. Therefore, the
$R_l(0)$ are finite for all $l\ge0$, as one usually requires
for physical solutions. In contrast,
the $Y_\sigma$ solutions would lead to  $R_l(r) \sim \r^{-l-1}$.
This shows that the choice $R_l(\r)\sim J_\sigma(z)$ gives the correct
physical solutions.

For $\r \rai$ the above solutions behave as
\be
R_l(\r) \sim \r^{\frac{\mu-1}{2}} \cos(\frac{1}{\r^\mu}-const.)~.
\ee
Thus, the probability integral in the asymptotic region can be estimated
as follows:
\bea
\int_R^\infty |R_l(\r)|^2 \r^2 d\r &\simeq &
 \int_R^\infty \r^{\mu+1} \cos^2(\frac{g}{|\mu|\r^\mu}-const.) d\r\le
 \int_R^\infty \r^{\mu+1} d \r \nn \\
&=& - \frac{1}{\mu+2} R^{\mu+2}~, ~~~~~ \mu < -2~.
\eea
This result can also be understood as follows: In order for the
probability integral to converge at infinity,
the integrand must fall off stronger than $1/\r$, which is the limiting
behavior that gives a logarithmic divergence. Hence,
for $\mu+1 < -1$ or $\mu ~<~ -2$, we obtain convergence at infinity.

Since, our solutions are finite at the origin, we conclude that our
solutions \rf{gensol} are normalizable for $\mu <-2$ for
all $l\ge 0$. These are the surprising discrete, yet unbound, states.

\subsection*{A.3: Physical solutions for $\mu >0$}

The choice of physical solutions is much more complicated for  $\mu >0$.
In this case, to obtain the $\r \raz$ limit we must use the
$z\rai $ limit in Eq.  \rf{asm5}. The solution \rf{gensol} behaves as
\be
R_l(\r) \sim \r^{(\mu-1)/2} \cos (g/(|\mu|\r^\mu) -const.)~, ~~~~ \r \raz
{}~.\lb{asz2}
\ee
$R_l(\r)$ is a rapidly oscillating function of $\r$, which is bounded by
$ \r^{(\mu-1)/2}$. This bound stays finite for $\mu \ge 1$, but goes to
infinity as $\r \raz$ for $ 1> \mu >0 $.
{\em In both cases, $R_l(\r)$ is
not an analytic function of $\r$ at the origin.}
This is not a customary
behavior at the origin.
However, it cannot be avoided
since both solutions, $\psi \propto J_\sigma(z)$ and $\psi \propto
Y_\sigma(z)$,
$\sigma\equiv(l+1/2)/|\mu|$, have the same type of asymptotic behavior
at infinite argument $z\equiv g/(|\mu|\r^\mu)$.

However, recalling that our potentials for $\mu >0$ are
unusually singular, we can relax the condition that
 $R_l(0)$ be finite.
{\em Instead,  we require  that
the probability of
finding the particle in any finite neighborhood of the origin should
be finite}:
\bea
\int_0^R |R_l(\r)|^2 \r^2 d\r
&=& \int_0^R \r^{\mu+1} \cos^2(\frac{g}{|\mu|\r^\mu}-const.) d\r\le
 \int_0^R \r^{\mu+1} d \r \nn \\
&=& \frac{1}{\mu+2} R^{\mu+2} < \infty~,
{}~~~~ \mu > 0~.
\eea
This shows that the probability is finite at the origin
for all positive $\mu$,  including $1> \mu >0$.

But the $\r \raz $ behavior for $\mu>0$ still does not enable us
to choose
between the $J_\sigma$ and the $Y_\sigma$ solutions. Both choices, or any
linear combinations of them, yield  physically acceptable solutions at the
origin.  However,
the $\r \rai$ limit can settle the matter. In this limit, we have
\be
R_l\sim J_\sigma(z)/\sqrt{\r}\sim \r^{-(l+1)}~, ~~~~ \r \rai ~, \lb{as4}
\ee
whereas
\be
R_l\sim Y_\sigma(z)/\sqrt{\r}\sim \r^{l}~,  ~~~~ \r \rai ~.
\ee
The asymptotic behavior \rf{as4} shows that our choice
\rf{gensol} leads to decaying solutions for $\r \rai$. In
fact, this  choice for  $\mu>0$ insures that
the whole probability integrated up to $\r \rai$ remains finite, and thus
these solutions correspond to bound states.

%*******************************************************

\section*{Appendix B: A Note on the Power-Law Potentials with $\gamma <0$}

Throughout the present paper, and in paper I \cite{db1},
{\em we have discussed the power-law
potentials only for $\gamma>0$ }.  For completeness, we add a
few comments on the $\gamma <0$ case.  There the power-law
potentials are
everywhere positive, $ V(r) >0$.  Therefore, they do not have any
classical solutions with $E=0$, since then the kinetic energy would
have to be negative.

However, quantum mechanically the $E=0$
situation is a little more interesting.  From Eq. \rf{gensol},
our solutions would now be of the form
\be
\hat{R}_l(\r) = \frac{1}{\r^{1/2}}
           {\mbox{\huge{J}}}
           _{\left(\frac{l+1/2}{|\mu|}\right)}
           \left(\frac{ig}
           {|\mu| \r^{\mu}}\right)
          \sim \frac{1}{\r^{1/2}}
           {\mbox{\huge{I}}}
           _{\left(\frac{l+1/2}{|\mu|}\right)}
           \left(\frac{g}
           {|\mu| \r^{\mu}}\right)~, ~~~ \mu \neq 0~,   \lb{last}
\ee
where  $ g \equiv |\sqrt{\gamma}|$  and  the $I$'s are the modified
Bessel functions \cite{modbes}.  The general solutions would also
involve the modified Bessel functions of the second kind, the $K$'s,
which can be written as linear combinations of the $I$'s.

We now can distinguish between two index regions:

(1) For $\nu >0$ or $\mu > -1$, the negative-$\gamma$
 potentials are positive definite and
repulsive from the origin. Hence, we expect
 scattering solutions.

(2) For $\nu <0$ or $\mu  < -1$, the negative-$\gamma$
potentials are confining potentials, which
can only have discrete bound states.
The corresponding energies must all be positive, similar to the energy
levels of the spherical harmonic oscillator.
Therefore, the solution of Eq. \rf{last} is not physical.

In this connection, and although not directly related to the nonexisting
$E=0$ physical states, we find it interesting to note that
Grosche and Steiner \cite{gs} were able to calculate the $E=0$
propagator for these $\nu <0$, negative-$\gamma$ potentials, by
using path integrals. They used this  propagator
to obtain interesting sum rules.

%********************************************************


\begin{thebibliography}{44}
\bibitem{ND} As a final note we
observe that the exceptional  $\nu = 2$ solutions of
Section 5.1 will also change their natures
as a function of $D$.
\end{thebibliography}
\end{document}